\documentstyle[12pt]{article}
\begin{document}

\vskip 3cm

\begin{center}

{\bf {A new differential calculus on noncommutative spaces}}\\

\vskip 2cm

{\bf R. P. Malik}
\footnote{ E-mail address: malik@boson.bose.res.in}\\
{\it S. N. Bose National Centre for Basic Sciences,}\\
{\it Block-JD, Sector-III, Salt Lake, Calcutta-700 091, India.}\\

\vskip .7cm

{\bf A. K. Mishra} 
\footnote{ E-mail address: mishra@imsc.ernet.in}
and
{\bf G. Rajasekaran}
\footnote{ E-mail address: graj@imsc.ernet.in}\\
{\it The Institute of Mathematical Sciences,}\\
{\it C.I.T Campus, Madras-600 113, India.}\\

\vskip 1.7cm

{\bf Abstract}

\end{center}

\noindent 
We develop a  $GL_{qp}(2)$ invariant differential calculus on a 
two-dimensional noncommutative quantum space. 
Here the co-ordinate space for the exterior quantum plane
is spanned by the differentials that are commutative 
(bosonic) in nature.

\baselineskip=24pt

The quantum group $GL_{qp}(n)$ covariant 
differential calculus on a quantum hyperplane is generated by three 
elements : (i) $\{x_i\} (i = 1, 2, 3, ......n)$, the noncommuting
co-ordinates of the quantum plane, 
(ii) differentials $\{dx_i\}$, which are
co-ordinates of the exterior quantum plane, and (iii) derivatives on the 
quantum plane, namely
$\{\frac{\partial}{\partial x_i}\}$ [1-4]. In addition, one also 
introduces an exterior differential $d$ [1]
$$
d \ = \ \sum_i dx_i \frac{\partial}{\partial x _i}. \eqno(1)
$$

In the classical limit, two distinct kinds of $dx$, satisfying either 
the algebraic relation

$$
\tilde{d} x_i \tilde{d} x_j 
+ \tilde{d} x_j \tilde{d} x_i \ = \ 0 \ \eqno(2)
$$
or the relation
$$
dx_i dx_j \ - \ dx_j dx_i \ = \ 0 \eqno(3)
$$

\noindent exist. The one-form $\tilde{d} x$, with $(\tilde{d}x)^2 = 0$, is 
introduced in the
context of differential geometry [5]. On the other hand,  $dx$ denotes the 
infinitesimal whose square is not
identically zero. The co-ordinate space for the exterior plane, 
spanned by $\{\tilde{d}x_i\}$,
is anticommutative (fermionic), and the one spanned by $\{dx_i\}$ is 
commutative (bosonic).

Much progress has been made in the area of 
noncommutative differential calculus (NDC) based on
a generalized $\tilde{d}x$ [1-4,6,7]. Is it possible to provide a consistent  
noncommutative differential calculus 
involving appropriately modified $dx$?  The answer to this query is in 
affirmative.
In the present paper, we develop the basic structure of this new calculus 
with a view to
evaluate the derivatives of a function having as its arguments  the 
co-ordinates of quantum plane
and/or exterior quantum plane (i.e., differentials), or any 
product of such functions.

We denote henceforth the conventional noncommutative differential calculus 
having $\tilde{d}
x_i$ as one of its elements as NDC-I, and the new differential calculus 
involving $dx_i$ as
NDC-II. The $d$ and $dx_i$ in eq.(1) now belong to NDC-II, 
whereas these quantities are to be
replaced by $\tilde{d}$ and $\tilde{d}x_i$ in NDC-I. It can be shown 
that $\tilde{d}^2 = 0$ [1],
though no such constraint exists for $d^2$. In fact, $d^n \ne 0$ for any 
integer $n =
0,1,2,\ldots$. Consequently, we have non-zero $d^n x_i$ - 
the n$^{th}$ differential in NDC-II,
with $d^{0} x_i \equiv x_i$. In the NDC-I, all higher order differentials, 
but for the one
corresponding to $n=1$, are zero. Here again, 
$\tilde{d}^{0} x_i \equiv x_i$.

The derivatives $\{ \frac{\partial}{\partial (\tilde{d} x_i)}\}$ on 
the exterior quantum plane,
though not defined originally, has been subsequently introduced [8]. 
We presently introduce the
analogous derivatives $\{\frac{\partial}{\partial (d^n x_i)}\}$. How 
these additional derivatives
modify $d$ in eq.(1) will be considered shortly.

We consider here a two dimensional system. This enables us to 
describe the basic structure of
the new calculus  without getting involved in the mathematical 
complexities when dimension exceeds two. The
case of higher dimensions will be considered in a subsequent paper.
We identify $d^n x_1 \equiv d^n x$ and $d^n x_2 \equiv d^n y$. In order to 
provide the details
of NDC-II, the first task is to determine the generalized commutation 
relations between $(d^n x,
d^m x), (d^n y, d^m y)$ and $(d^n x, d^m y)$ for arbitrary positive 
integers $n$ and $m$. In addition, either or both
$n$ and $m$ can be zero also. To achieve this, one needs only the 
relations corresponding to $m =
n,n+1$. The remaining relations follow by the successive application 
of the exterior differential $d$.
The $d$  satisfies the Leibnitz rule
$$
d\;(fg) = (df)\;g + f\;(dg) \eqno(4)
$$
with $f$ and $g$ being functions of $(d^nx, d^m y)$.

We begin with the following general relations
$$
(d^n x) (d^n y)  =  C_1 (d^n y) (d^n x) \eqno(5)
$$
$$
(d^n x) (d^{n+1} x)  =  C_2 (d^{n+1} x) (d^n x) \eqno(6)
$$
$$
(d^n y) (d^{n+1} y)  =  C_3 (d^{n+1} y) (d^n y) \eqno(7)
$$
$$
(d^n y) (d^{n+1} x)  =  C_4 (d^{n+1} x) (d^n y). \eqno(8)
$$
The various $C^{\prime s }$ are non-zero constants. Applying $d$ on 
relation (5) and using eq.(8), we obtain the last relation
$$
(d^n x) (d^{n+1} y) \ = \ C_1 (d^{n+1} y) (d^n x) 
+ (C_1 C_4-1) (d^{n+1} x) (d^n y). \eqno(9)
$$
The requirement of associativity and the validity of 
Yang-Baxter equation demands that when $d^k x, d^k
y~ (k = n, n+1)$ are commuted through eqs. (5-8), either from the 
left or from the right, no new relation should 
emerge [9]. This condition is satisfied provided
$$
C_2 \ = \ C_3 \ = C_1 C_4. \eqno(10)
$$

It can be seen that when $C_1~ = ~ q$ and $C_4~=~ p$, 
the relations (5-8) are invariant
under the quantum group transformations 
$$
\pmatrix {d^n x \cr d^n y} \rightarrow  
\pmatrix {d^n x^{\prime} \cr d^n y^{\prime}} =
\pmatrix { A & B \cr C & D} \pmatrix {d^n x \cr d^n y}  \eqno(11)
$$
where $A, B, C, D$ are the elements of a 
$2 \times 2$ $GL_{qp}(2)$ matrix obeying the $q$-algebraic relations 
in rows and columns  as 
$$ \begin{array}{lclcl}
AB = pBA & , & AC = qCA & , & BC 
= \frac{\displaystyle q}{\displaystyle p} CB \\
CD = pDC & , & BD = qDB & , & AD-DA = (p-q^{-1}) BC  \end{array} \eqno(12)
$$
and these elements are assumed to commute with all the differentials 
in eqs.(5-8).

Substituing the values of C's in eqs. (5-9), we get 
$$
\xi^n \eta^n \ = \ q \eta^n \xi ^n \eqno(13)
$$
$$
\xi^n \xi^{n+1} \ = \ pq \xi^{n+1} \xi^n \eqno(14)
$$
$$
\eta^n \eta^{n+1} \ = \ pq \eta^{n+1} \eta^n \eqno(15)
$$
$$
\eta^n \xi^{n+1} \ = \ p \xi^{n+1} \eta^n \eqno(16)
$$
$$
\xi^n \eta^{n+1} \ = \ q\eta^{n+1} \xi^n + (pq-1) \xi^{n+1} \eta^n \eqno(17)
$$
where for the notational convenience, we have introduced
$$
d^k x \ \equiv \xi^k \quad ; \quad d^k y \ \equiv \eta^k. \eqno(18)
$$
We note here that superscripts  of $\xi$ and $\eta$ 
denote an index, and not the exponent. 

Applying the $d$ consecutively on relations (13-17), and after some 
algebraic manipulations, the
following generalizations are obtained for equations (14-17) 
corresponding to any $m > n$:
$$
\xi^n \xi^m \ = \ pq \xi^m \xi^n + (pq-1) 
\sum^{m-n-1}_{\stackrel{r=1}{m-n-r >0}} \xi^{n+r}
\xi^{m-r} \eqno(19)
$$
$$
\eta^n \eta^m \ = \ pq \eta^m \eta^n 
+ (pq-1) \sum^{m-n-1}_{\stackrel{r=1}{m-n-r >0}} \eta^{n+r}
\eta^{m-r} \eqno(20)
$$
$$
\eta^n \xi^m \ = \ p \xi^m \eta^n 
+ (pq-1) \sum^{m-n-1}_{\stackrel{r=1}{m-n-r >0}} 
\eta^{m-r} \xi^{n+r} \eqno(21)
$$
$$
\xi^n \eta^m \ = \ q \eta^m \xi^n 
+ (pq-1) \sum^{m-n-1}_{\stackrel{r=0}{m-n-r >0}} \xi^{m-r} 
\eta^{n+r}. \eqno(22)
$$
Equations (13,19-22) provide all possible commutation relations 
between the differentials.

The ultimate aim of the present analysis is to provide relations 
which enable one to evaluate the
derivatives of functions like $f$ and $g$ introduced in eq.4. Such 
derivatives can be obtained using the
commutation relations  between  $( \xi^n, \eta^m)$  and derivatives 
 $( \frac{\partial}{\partial \xi^k}$, 
 $\frac{\partial}{\partial \eta^l})$. To
obtain these relations, we need explicit form of the exterior differential $d$. Accordingly, we
generalize the expression (1) as
$$
d \ = \ \sum^{\infty}_{n=1} \left( \xi^n \frac{\partial}{\partial \xi^{n-1}} 
+ \eta^n
\frac{\partial}{\partial \eta^{n-1}}\right). \eqno(23)
$$
Applying $d$ on $\xi^n f$ and using (23), we obtain
$$
d(\xi^n f) \ = \ \sum_{m >0} \xi^m \frac{\partial}
{\partial \xi ^{m-1}} \xi^n f + \sum_{m > 0} \eta^m
\frac{\partial}{\partial \eta^{m-1}} \xi^n f. \eqno(24)
$$
The Leibnitz rule (cf. eq.4) leads to 
$$
d\;(\xi^n f) \ = \ \xi^{n+1}\; f + \xi^n \;d\;f \eqno(25)
$$
where $\xi^{n+1} \ = \ d\xi^n \ = \ d(d^n x) \ = \ d^{n+1} x$.
Substituting $d$ from eq.(23) in the
right hand side of eq.(25), we get 
$$
d(\xi^n f) \ = \ \xi^{n+1} f 
+ \xi^n \sum_{m > 0} \xi^m \frac{\partial}{\partial \xi^{m-1}} f + \xi^n
\sum_{m >0} \eta^m \frac{\partial}{\partial \eta^{m-1}} f. \eqno(26)
$$

Next, we use relations (13,19-22) to exchange the order of 
$\xi^n \xi^m$ and $\xi^n \eta^m$ in eq.(26).
Thereafter, comparing the coefficients of $\xi^m$ and $\eta^m$ 
in the resulting expression with the coefficients of
$\xi^m$ and $\eta^m$ in eq.(24), we obtain the following set of required 
commutation relations
$$
\frac{\partial}{\partial \xi^m} \xi^n \ 
= \ \frac{1}{pq} \xi^n \frac{\partial}{\partial \xi^m} -
\frac{(pq-1)}{pq} \sum^m_{k=1} \xi^{n-k} \frac{\partial}{\partial 
\xi^{m-k}}, \quad  0 \le m \le n-2 \eqno(27)
$$
$$
\frac{\partial}{\partial \xi^m} \xi^n \ 
= \ \xi^n \frac{\partial}{\partial \xi^m}, \quad  \ m = n-1 \eqno(28)
$$

$$
\frac{\partial}{\partial \xi^m} \xi^n \ 
= \delta_{n,m} + \ pq \xi^n \frac{\partial}{\partial \xi^m} + (pq-1)
\sum^\infty_{k=1} \left( \xi^{n+k} \frac{\partial}{\partial 
\xi^{m+k}} + \eta^{n+k-1}
\frac{\partial}{\partial \eta^{m+k-1}}\right),\; 
m > n-1 \eqno(29)
$$
$$
\frac{\partial}{\partial \eta^m} \xi^n \ 
= \ \frac{1}{p} \xi^n \frac{\partial}{\partial \eta^m} -
\frac{(pq-1)}{p} \sum^m_{k=1} \xi^{n-k} 
\frac{\partial}{\partial \eta^{m-k}}, 
\quad  0 \le m \le n -2 \eqno(30)
$$
$$
\frac{\partial}{\partial \eta^m} \xi^n \ 
= \ q\xi^n \frac{\partial }{\partial \eta^m}, \quad m
\ge n-1. \eqno(31)
$$
Repeating the same procedure with $d (\eta f)$ in the place of 
$d (\xi f)$,  we get
$$
\frac{\partial}{\partial \eta^m} \eta^n \ 
= \ \frac{1}{pq} \eta^n \frac{\partial}{\partial \eta^m} -
\frac{(pq-1)}{pq} \sum^m_{k=1} \eta^{n-k} 
\frac{\partial}{\partial \eta^{m-k}}, \quad   0 \le m \le n-2
\eqno(32)
$$
$$
\frac{\partial}{\partial \eta^m} \eta^n \ 
= \ \eta^n \frac{\partial }{\partial \eta^m},  \quad   m = n-1
\eqno(33)
$$
$$
\frac{\partial}{\partial \eta^m} \eta^n \ 
= \delta_{n,m} + \ {pq} \eta^n \frac{\partial}{\partial \eta^m} +
(pq-1) \sum^\infty_{k=1} \left(\xi^{n+k} 
\frac{\partial}{\partial \xi^{m+k}} \ + \eta^{m+k} 
\frac{\partial}{\partial \eta^{m+k}}\right),  
\quad   m > n-1 \eqno(34)
$$
$$
\frac{\partial}{\partial \xi^m} \eta^n \ 
= \ \frac{1}{q} \eta^n \frac{\partial }{\partial \xi^m} -
\frac{(pq-1)}{q} \sum^m_{k=1} \eta^{n-k} 
\frac{\partial}{\partial \xi^{m-k}}, \quad  0 \le m \le n-2 
\eqno(35)
$$
$$
\frac{\partial}{\partial \xi^m} \eta^n \ 
= \ \frac{1}{q} \eta^n \frac{\partial }{\partial \xi^m} -
\frac{(pq-1)}{q} \sum^{n-1}_{k=1} \eta^{k} 
\frac{\partial}{\partial \xi^{k-1}}, \quad  m = n-1 
\eqno(36)
$$
$$
\frac{\partial}{\partial \xi^m} \eta^n \ 
= \ p \eta^n \frac{\partial}{\partial \xi^m}, \quad m >
n-1. \eqno(37)
$$

Given any function of $(\xi^n, \eta^m)$,  the 
relations (27-37) enable one to evaluate 
its derivative as well as any sequence of  
derivatives with respect to $\xi^k$ and  $\eta^l$.

We may note the importance of the new calculus in the context of 
particle dynamics 
on a noncommutative space. For example, it is the $dx$ and $d^2x$ 
which are related to the
velocity $\dot{x}$ and acceleration $\ddot{x}$ through 
the relations $dx = \dot{x} dt$ and $d^2 x =
\ddot{x}(dt)^2$, where the time $t$ is taken to be a 
commuting number. On the other hand, a
relation like $\tilde{d} x = \dot{x} dt$ would make the 
velocity nilpotent, and the acceleration
can not be related to $\tilde{d}^2 x$ as this quantity is identically zero.

Is it possible to provide a connection between NDC-I and the new  
noncommutative calculus?
Can the new calculus be mapped to a Fock-space structure? These queries 
along with the 
possible applications will be considered in our future communications.

\vskip .5cm

\newpage

\baselineskip = 12pt
\noindent {\bf References}

\begin{enumerate}
\item Wess J and Zumino B  1990
{\it  Nucl. Phys. (Proc. Suppl.)} {\bf B18} 302 
\item Manin Yu I 1989 {\it  Comm. Math. Phys.} {\bf 123} 163
\item Woronowicz S L 1989 {\it Comm. Math. Phys.} {\bf 122} 125
\item Schirrmacher A 1991 {\it  Z. Phys.}  {\bf C50} 321    
\item Spivak M 1970 {\it  A comprehensive Introduction to Differential
 Geometry} (Publish or Perish, Boston) Vol. {\bf 1}
\item Pusz W and Woronowicz S L 1989 {\it  Rep. Math. Phys.} {\bf 27} 231
\item Soni S K 1991 {\it J. Phys.}  {\bf A24} L169 
\item Mishra A K and Rajasekaran G. 1997
      {\it  J. Math. Phys.} {\bf 38} 466
\item Lukin M, Stern A and Yakushin I  1993
      {\it J. Phys.}  {\bf A26} 5115
\end{enumerate}

\end{document}